\documentclass[useAMS,usegraphicx,usenatbib]{mn2e}

\usepackage{times}
\usepackage{amssymb}
\usepackage{amsmath}
\usepackage{rotate}
\newif\ifAMStwofonts
\AMStwofontstrue

\def\dof{{$dof$}}
\def\cloudy{\textsc{cloudy}}
\def\xmm{{\it XMM-Newton}}
\def\chandra{{\it Chandra}}
\def\sax{{\it BeppoSAX}}
\def\mcg{{MCG--6-30-15}}
\def\mrk766{{Mrk 766}}

\def\cm{{\rm\thinspace cm}}
\def\ps{{\rm\thinspace s^{-1}}}
\def\km{{\rm\thinspace km}}
\def\s{{\rm\thinspace s}}
\def\pscm{\hbox{$\cm^{-2}\,$}}
\def\erg{{\rm\thinspace erg}}
\def\ergps{\hbox{$\erg\s^{-1}\,$}}
\def\kmps{\hbox{$\km\ps\,$}} 
\def\Mpc{{\rm\thinspace Mpc}}
\def\kmpspMpc{\hbox{$\kmps\Mpc^{-1}\,$}}

\title[A softer look at \mcg\ with \xmm]{A softer look at \mcg\ with
  \xmm}

\author[A. K. Turner et al.]  {\parbox[]{6.in} {A. K.
    Turner$^{1}$\thanks{E-mail: akt21@ast.cam.ac.uk}, A. C.
    Fabian$^{1}$, S. Vaughan$^{1}$ and J. C. Lee$^{2}$
    \\
    \footnotesize
    $^{1}$Institute of Astronomy, Madingley Road, Cambridge CB3 0HA\\
    $^{2}$Chandra Fellow, Massachusetts Institute of Technology, Center for Space Research, 77 Massachusetts Ave. NE80, Cambridge, MA 02139, USA\\
  }}

\begin{document}

\maketitle

\label{firstpage}

\begin{abstract}
  We present the analysis and first results from the Reflection
  Grating Spectrometer (RGS) during the 320~ks \xmm\ observation of
  the Seyfert 1 galaxy \mcg. The spectrum is marked by a sharp drop in
  flux at 0.7~keV which has been interpreted by
  \citeauthor{brand-ray01} using RGS spectra from an earlier and
  shorter observation as the blue wing of a strong relativistic OVIII
  emission line and by \citeauthor{lee01} using a \chandra\ spectrum
  as due to a dusty warm absorber. We find that the drop is well
  explained by the FeI L$_{2,3}$ absorption edges and obtain
  reasonable fits over the 0.32--1.7~keV band using a multizone, dusty
  warm absorber model constructed using the photoionization code
  \cloudy. Some residuals remain which could be due to emission from a
  relativistic disc, but at a much weaker level than from any simple
  model relying on relativistic emission lines alone. A model based on
  such emission lines can be made to fit if sufficient (warm)
  absorption is added, although the line strengths exceed those
  expected. In order to distinguish further whether the spectral shape
  is dominated by absorption or emission, we examined the difference
  spectrum between the highest and lowest flux states of the source.
  The EPIC pn data indicate that this is a power-law in the 3--10~keV
  band which, if extrapolated to lower energies, reveals the
  absorption function acting on the intrinsic spectrum, provided that
  any emission lines do not scale exactly with the continuum. We find
  that this function matches our dusty warm absorber model well if the
  power-law steepens below 2~keV. The soft X-ray spectrum is therefore
  dominated by absorption structures, with the equivalent width of any
  individual emission lines in the residuals being below about 30~eV.
\end{abstract}


\section{Introduction}

\mcg\ is one of the most studied Seyfert 1 galaxies in the X-ray band
due to its high X-ray flux and relative closeness (z = 0.00775,
$L_{2-10~\mathrm{keV}}\!\sim\!5\times10^{42}$\ergps; $H_0=70$\kmpspMpc
). At energies above 2~keV the X-ray spectrum consists of a power-law
with $\Gamma \simeq 1.9$, skewed iron K$\alpha$ line and reflection
hump \citep[e.g.][]{fabian02}. At softer energies much
controversy still exist concerning the observed spectral features.
Before the use of grating spectra, sharp drops in flux observed in
soft X-ray spectra were interpreted as absorption edges due to OVII
and OVIII in ionized gas within the AGN environment \citep[``the warm
absorber'';][]{otani96,reynolds97sample,george98}.

However the increase in spectral resolution afforded by the use of
grating spectrometers onboard \xmm\ and \chandra\ has challenged the
view that reprocessing of soft X-rays by photoionized gas is
responsible for the observed spectral features. \citet{brand-ray01}
proposed that sharp drops seen in the Reflection Grating Spectrometer
(RGS) spectra from \xmm\ observations of \mcg\ and \mrk766\ at
$\sim\!0.39$, $\sim\!0.55$ and $\sim\!0.7$~keV were due to the blue
wings of relativistically blurred emission lines of CVI, NVII and
OVIII Ly$\alpha$, the blurring mechanism being the same as that
  used to explain the skewness of the Fe K$\alpha$ line. The reason
for adopting this view was that the position of the sharp drop at
$\sim0.7$~keV, if caused by an OVII absorption edge inferred an
unrealistically high infall velocity of matter of $\sim16,000$~\kmps.
\citet{lee01} countered this suggestion with \chandra\ High Energy
Transmission Grating Spectrometer (HETGS) observations in which the
16,000~\kmps discrepancy in position of the OVII edge was explained by
the blending together of resonance absorption lines red-ward of
the edge and by L-shell absorption from neutral iron which is probably
in the form of dust (the ``dusty warm absorber'' model; DWA).  In a
re-analysis of the original RGS data \citet{sako01} fitted both the
DWA and blurred emission line models to the original RGS data.  These
workers argue that the models of \citet{lee01} over-predicted the flux
below the passband of the HETG instrument and that the relativistic
line interpretation provides a better fit to the data than the DWA
interpretation.

Given the evidence for a relativistically-blurred iron line and
reflection continuum in \mcg, some relativistically-blurred soft X-ray
lines are also expected in the spectrum
\citep*{ross93,nayakshin00,rozanska02} The strength of these lines is
however predicted to be much smaller, and the blue wing broader, than
those claimed from the RGS spectra \citep*{ballantyne02}. The
determination of the underlying soft X-ray continuum is of importance
for understanding the X-ray irradiation of accretion discs.

This paper presents the RGS data for the long $\sim\!320$~ks
observation of \mcg\ \citep{fabian02}. In particular we test to see if
a DWA or relativistically-blurred emission line model can explain the
overall spectral shape and see what can be inferred about the
underlying continuum. The rest of the paper is organised as follows.
Section \ref{data_reduction} describes the data reduction. In Section
\ref{simple_absorption_model} absorption across the 0.7~keV drop is
examined.  Section \ref{photoionization_model} presents a
photoionization model of the warm absorber used in Section
\ref{spectral_fitting}, where DWA and the relativistically-blurred
emission line models are fit to the data. In Section
\ref{empirical_model} an attempt is made to use spectral variability
to determine the underlying continuum. The results are discussed in
Section \ref{discussion}. This paper does not represent a full
description of the data but a first step in modelling the RGS spectrum
of the DWA in \mcg\ and a first look at some of the key features in
the data. Plasma diagnostics of individual species will be presented
in a later paper.

\section{Data Reduction}

\label{data_reduction}

\mcg\ was observed by \xmm\ on revolutions 301, 302 and 303 (2001 July
31 -- 2001 August 5) for 320 ks \citep{fabian02}. The observation data
files (ODFs) were reduced with version 5.3.3 of the \xmm\ Science
Analysis Software ({\sc sas}) using the standard processing chains and
the calibrations from 2002 December. The final few ks of data from
each revolution showed strong background flaring, these periods were
removed from the data extraction. The data were grouped so that at
least 20 counts were in each bin and were fitted to trial models in
{\sc xspec v11.2} \citep{arnaud96}. Data above 1.7~keV were ignored
due to the known calibration uncertainties at the mirror Au edge.
Unless otherwise stated errors quoted are for 90 per cent confidence
level for one interesting parameter ($\Delta\chi^2 = 2.7$). The data
are plotted in the observation frame throughout.

\section{Simple absorption model}

\label{simple_absorption_model}

\begin{figure}
\rotatebox{270}{
\resizebox{!}{\columnwidth}
{\includegraphics{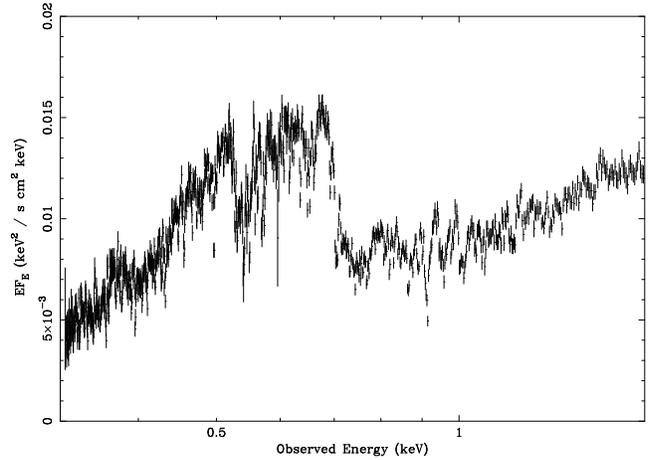}}}
\caption{
  Combined and fluxed RGS data from all three satellite revolutions.
}
\label{fig:flux_all_orbits.eps}
\end{figure}

The fluxed RGS spectrum from all three revolutions is shown in
Fig.~\ref{fig:flux_all_orbits.eps}.
Evident in the spectra are the drops in flux at 0.39, 0.54 and
0.7~keV. The high spectral resolution of the RGS requires the
inclusion of individual resonance absorption lines when considering
the effect of photoionized absorption. Specifically, at the red side of
an absorption edge there is a series of resonant absorption lines that
converge to the edge energy. These have the effect of reducing the
observed edge energy when convolved with intrinsic Doppler broadening
and the spectral resolution of the instrument.

\begin{figure}
  \rotatebox{270}{ \resizebox{!}{\columnwidth}
    {\includegraphics{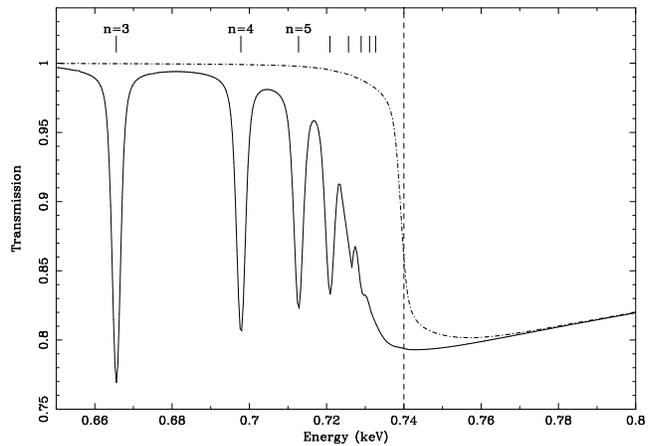}}}
\caption{
  Detail of the predicted absorption near the OVII K-edge convolved
  with the RGS resolution for an OVII column density of
  $10^{18}$~cm$^{-2}$ and turbulent velocity of 100~\kmps.
  (\emph{solid}): Absorption including photoelectric edge and
  $1s^2-1snp$ resonance lines, (\emph{dot-dash}): Absorption including
  photoelectric edge only, (\emph{dash}): Position of OVII K-edge.}
\label{fig:OVII_edge}
\end{figure}

Models of the cross-section of OVII and OVIII near their K-edges were
constructed using the continuum cross-section fitting formulae of
\citet{verner95}, the resonance absorption line data of
\citet{verner96} and the fitting formulae for line absorption series
near absorption edges of \citet{mewe77}. The transmission through a
column density of $10^{18}$~cm$^{-2}$ of OVII, calculated assuming a
turbulent velocity (b parameter) of 100~\kmps, is plotted in
Fig.~\ref{fig:OVII_edge}. This shows the effect of the resonance lines
on the apparent position of the edge, which is clearly shifted to
lower energies. Also present at 0.7~keV are the FeI L-shell absorption
edges which have fine structure in their cross-section at the
L$_{2,3}$ edge as described by \citet{kortright00}.

\begin{figure}
\rotatebox{270}{
\resizebox{!}{\columnwidth}
{\includegraphics{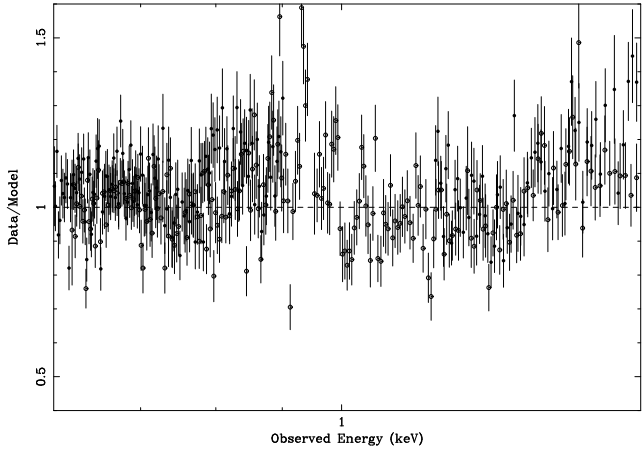}}}
\caption{
  Ratio of data for (\emph{filled circles}) RGS1 and (\emph{open
    circles}) RGS2 from orbit 301 to a model consisting of a
  power-law, Galactic absorption and OVII, OVIII and FeI absorption.}
\label{fig:ratio_036c}
\end{figure}

\begin{figure}
\rotatebox{270}{
\resizebox{!}{\columnwidth}
{\includegraphics{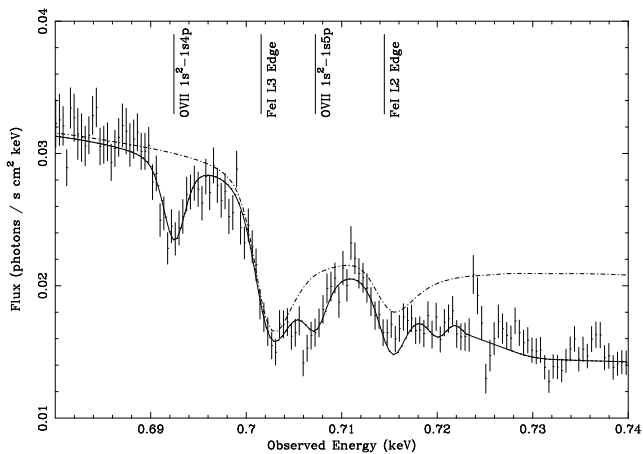}}}
\caption{
  Details of fluxed spectrum for all three orbits compared to a model
  of OVII, OVIII and FeI absorption (\emph{solid}) and FeI absorption
  only (\emph{dash-dot}).}
\label{fig:FeEdge_detail_4}
\end{figure}

Fig.~\ref{fig:ratio_036c} shows the revolution 301 data fitted to a
power-law, Galactic absorption \citep*[$N_{\rm H}=4.06 \times
10^{20}$~cm$^{-2}$;][]{elvis89} and absorption by OVII, OVIII and FeI
across the energy range 0.6 -- 1.7~keV. The best fit ($\chi^2 =
9003.9$ for 6625 degrees of freedom (\dof)) requires a photon index of
$\Gamma = 2.36\pm0.02$ and column densities of $\log N_{\rm OVII} =
18.22\pm0.02$, $\log N_{\rm OVIII} = 18.47\pm0.02$ and $\log N_{\rm
  FeI} = 17.29\pm0.02$.  Although the best fitting model gives a poor
overall $\chi^2_\nu$, it reproduces the detailed shape of the drop at
$\sim$0.7~keV remarkably well as can be seen from
Fig.~\ref{fig:FeEdge_detail_4} where the model is compared to the
fluxed spectrum from all three orbits. The sharp drops at 0.7 and
0.713~keV are clearly fitted by the Fe I absorption profile. Also
evident are the resonant absorption lines of OVII. The FeI L$_3$ edge
is observed at $0.7023\pm0.0007$~keV which corresponds to an velocity
of $-320\pm340$\kmps with respect to the rest frame of the source if
the iron is in the metallic form described by \citet{kortright00}. The
absolute error on the energy scale of measurements of the
cross-section of iron oxides are required before it can be determined
whether the observed iron is in the metallic or iron oxide form. The
derived power-law photon index is higher than the photon index of
$\sim1.9$ derived with the EPIC instrument above 2.5~keV
\citep{fabian02} suggesting the emission below 1--2~keV has a soft
excess relative to an extrapolation of the best fitting continuum
above 2~keV \citep[see also][]{lee01}.  The residuals around 1~keV are
due to absorption lines of FeXIII-XXIV and NeIX-X.

\section{Photoionization model}

\label{photoionization_model}

While the previous Section suggests that absorption by FeI, OVII and
OVIII can explain the sharp drop in flux at 0.7 keV, it does not
provide a self-consistent model of absorption from a photoionized
absorber. In order to explore the properties of any warm absorption in
\mcg\ and provide a model of this absorption the {\tt 96beta5} version
of the photoionization code \cloudy\ was used to calculate the
absorbed spectrum produced by the passage of a continuum through a
layer of gas \citep{ferland98}. This version of \cloudy\ was used
since it includes the atomic transitions of the iron Unresolved
Transition Array \citep*[UTA;][]{behar01} which have been observed
previously in Seyfert 1 X-ray spectra. Above 1 keV the incident
continuum is taken to have the spectral shape inferred from fitting
the EPIC and \sax\ data \citep{fabian02}. Below 1.0 keV the spectrum
is assumed to be a power-law with $\Gamma = 2.0-3.0$ down to a break
energy of 10 eV and $\Gamma=-1.5$ below this. The normalization of the
incident spectrum is determined from the X-ray luminosity in the 2--10
keV range, $L_{2-10}$, inferred from the EPIC and \sax\ data.  Also
included in the incident continuum is a blackbody component with a
temperature of $2.5\times10^5$~K and bolometric luminosity of
$6\times10^{43}$~\ergps. The gas is assumed to have a constant density
and Solar elemental abundance with a turbulent velocity of 100\kmps.
This turbulent velocity is used since the lines appear unresolved in
the \chandra\ HETGS spectrum of \citet{lee01}.

The code was used to generate a grid of incident and absorbed spectra
as well as a list of absorption line optical depths at line centre
where greater than 0.01. The parameters that were varied across the
grid were column density of the gas ($N_\mathrm{H} = 10^{20} - 10^{22}
\mathrm{cm}^{-2}$), inner radius of cloud from central source ($R =
10^{16} - 10^{19} \mathrm{cm}$), ionization parameter of the cloud
($\xi = 10^{-5} - 10^{3}$) and spectral index between 10 eV and 1 keV
($\Gamma = 2.0 - 3.0$). The density of the gas, $n_\mathrm{H}$, is set
from $R$, $\xi$ and $L_{2-10}$, through the formula:
\[
\xi = \frac{L_{2-10}}{n_\mathrm{H}R^2}
\]
The resulting transmitted continua (i.e. including only
  bound-free edges) were divided by the incident continua to produce
a continuum absorption model. The absorption lines were then included
using a Gaussian opacity profile with a b width of 100 \kmps.

In using only the turbulent velocity as the line broadening mechanism
the natural and thermal broadening of the lines are neglected. Thermal
broadening is ignored since for the species of interest in a
photoionized gas thermal broadening velocities are only 10-20 per cent
of the turbulent velocity \citep*{nicastro99}. \citet{lee01} show that
various absorption lines from OVII and OVIII lie on the flat part of
the curve of growth for column densities of interest here, justifying
our neglect of natural line broadening.

\section{Spectral fitting}

\label{spectral_fitting}

\subsection{Dusty warm absorber model}

\label{DWA_model}

\begin{figure}
\rotatebox{270}{
\resizebox{!}{\columnwidth}
{\includegraphics{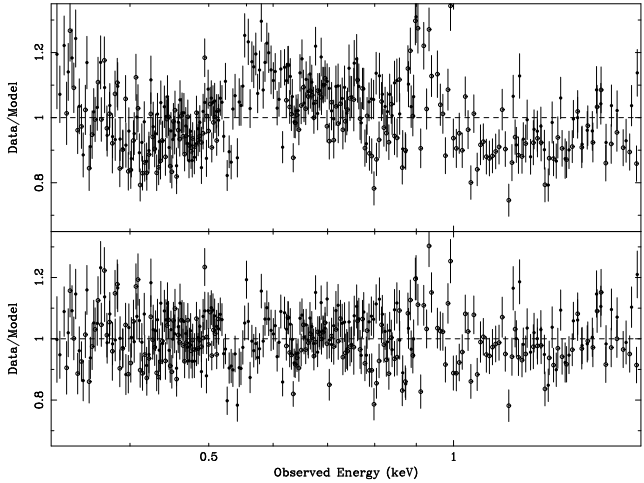}}}
\caption{
  Ratio of data from RGS1 (\emph{filled circles}) and RGS2 (\emph{open
    circles}) to the best fitting DWA models. Only data for orbit 301
  is shown here. (\emph{a}) Four zone DWA model with underlying broken
  power-law. (\emph{b}) Four zone DWA model with underlying broken
  power-law and excess broad emission at $\sim0.56$~keV.}
\label{fig:ratio_dwa}
\end{figure}

We now fit the warm absorber photoionization model to the data from
all three revolutions. Galactic absorption, FeI absorption and
multiple warm absorption zones are included to account for absorption
of the continuum. The absorption component is kept fixed between data
from different revolutions while the parameters of the underlying
continuum are allowed to vary. The continuum is taken to be a broken
power-law with a break energy of 1~keV, spectral index of 2.2 above
the break energy and a variable spectral index below the break energy.
The use of a broken power-law is justified since the variability
suggests some break in the continuum at $\sim1-2$~keV
\citep{fabian02,fabian03}. Fitting in the range 0.6--1.7~keV with a
three zone warm absorber gives $\chi^2 = 8412.6/6620$~\dof. The fit
parameters for this model are given in Table \ref{table_dwa} (model
1). Three zones are required since two distinct velocity components
are seen in the absorption lines \citep{sako01} and features from a
broad range of ionization parameters are present.  Extending the model
over the whole RGS passband reveals that the predicted model flux is
far higher than the observed data below 0.5~keV. The smooth nature of
the discrepancy suggests additional cold absorption is needed,
possibly in the form of absorption from the ISM within \mcg.  However,
including additional cold absorption in our fit to the 0.32--1.7~keV
energy range produces a large OI edge that is not seen in the data.
One possibility is that gas is partly ionized so most of the O is
ionized to higher species reducing the depth of the OI edge.  A four
zone warm absorber produces a fit with $\chi^2 = 16601.9/12053$~\dof,
where one of the zones has a low ionization parameter $\log\xi=-4.4$
accounting for the excess low energy absorption. Table \ref{table_dwa}
(model 2) gives the fit parameters for this model. The residuals for
this model are shown in Fig.~\ref{fig:ratio_dwa}a. As mentioned the
lowest ionization parameter zone is required to cause the spectrum to
turn down at low energies without too deep a neutral oxygen edge.  The
medium ionization parameter zone is required to fit the OVII edge and
lines, the drop in flux at $\sim0.39$~keV due to the CV edge and the
Fe UTA at 0.7--0.8~keV which takes the form of a trough in the data.
The higher parameter zones are required to fit a range of absorption
line features around 1~keV as well as any OVIII features.  The
majority of the OVII column density is located in zone 2
($\log\xi\sim-0.5$) while the majority of the OVIII column density is
located in the highest ionization parameter zones 3 and 4
($\log\xi\sim1.5$). The best fitting DWA model contains less OVII and
OVIII column density than the simple absorption model presented in
section \ref{simple_absorption_model} since other absorption
components are present in the photoionization models, such as the Fe
UTA, that can compensate for some of the OVII and OVIII column.
Although residuals are present they take the form of broad features,
whereas the region across the 0.7~keV drop is reasonably well fitted.
\begin{figure}
\rotatebox{270}{
\resizebox{!}{\columnwidth}
{\includegraphics{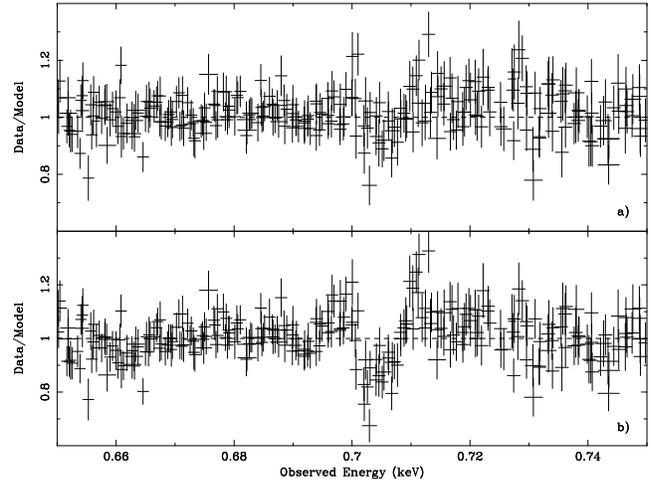}}}
\caption{
  (\emph{a}) Detail of residuals for the best fitting DWA model in the
  0.65--0.75~keV region. The six data sets are shown simultaneously
  (one from each revolution and instrument). (\emph{b}) Same as
  (\emph{a}) except for the best fitting REL model.}
\label{fig:ratio_detail}
\end{figure}

We now add a broad Gaussian emission component at $\sim0.6$~keV to
account for any excess emission in this region possibly from
reflection components. The best-fitting model produces a significant
improvement in the fit with $\chi^2 = 15977.3/12051$~\dof\ as can be
seen in Fig.~\ref{fig:ratio_dwa}b (see model 3 in Table
\ref{table_dwa} for the fit parameters of this model). The line has an
equivalent width of $\sim$120~eV and is very broad with
$\sigma\sim$170~eV.  The model fits the region around the 0.7~keV drop
extremely well (see Fig.~\ref{fig:ratio_detail}a). However, the broad
Gaussian shape is not that of a single relativistic disc line and
should be considered as only an approximation of the shape of the
residuals. Adding the broad Gaussian emission line results in changes
to the best fitting DWA component. The broad emission line has the
effect of fitting the excess in the residuals in the range
$\sim$0.5--0.9~keV.  This produces an excess below $\sim$0.4~keV and
above $\sim$1.35~keV. The DWA component compensates to fit these
excesses by, at low energy, increasing the ionisation parameter of the
lowest ionization parameter zone (zone 1) which results in absorption
that rolls over less at low energies and is flatter, and at high
energies by decreasing the column density of the high ionization
parameter zones which results in less absorption above $\sim1.35$~keV
flattening the absorption.

\begin{table*}
\centering
\caption{Results of fits of the DWA model to all three revolutions
  of RGS data. Line energy of the broad Gaussian emission
  line ($E_{\mathrm{g}}$) is given in units of keV. Width
  ($\sigma_{\mathrm{g}}$) and equivalent width ($EW_{\mathrm{g}}$) of
  the broad Gaussian emission line are given in eV. Column density
  ($N_{\mathrm{X}}$) is given in units of $\log$\pscm. Velocity ($v$)
  is given in \kmps.\label{table_dwa}}
\begin{tabular}{ccccc}                
\hline
model & zone & 1 & 2 & 3 \\
\hline\hline
$\Gamma_1$ & & $2.72\pm0.03$ & $2.91\pm0.02$ & $2.68\pm0.04$ \\
$E_{\mathrm{g}}$ & & -- & -- & $0.54\pm0.03$ \\
$\sigma_{\mathrm{g}}$ & & -- & -- & $176\pm15$ \\
$EW_{\mathrm{g}}$ & & -- & -- & $121\pm12$ \\
$N_{\mathrm{FeI}}$ & & $17.28\pm0.03$ & $17.22\pm0.02$ &
$17.25\pm0.03$ \\
\hline
$N_{\mathrm{H}}$ & 1 & $21.37\pm0.04$ & $20.86\pm0.01$ &
$21.12\pm0.04$ \\
$\log\xi$ & 1 & $-0.47\pm0.03$ & $-4.42\pm0.23$ & $-1.92\pm0.02$\\
$v$ & 1 & 0$^{f}$ & 0$^{f}$ & 0$^{f}$\\
\hline
$N_{\mathrm{H}}$ & 2 & $21.72\pm0.05$ & $21.44\pm0.03$ &
$21.44\pm0.03$ \\
$\log\xi$ & 2 & $1.43\pm0.03$ & $-0.52\pm0.02$ & $-0.36\pm0.03$ \\
$v$ & 2 & 0$^{f}$ & 0$^{f}$ & 0$^{f}$\\
\hline
$N_{\mathrm{H}}$ & 3 & 22.00$^{f}$ & 22.00$^{f}$ & $21.73\pm0.10$ \\
$\log\xi$ & 3 & $2.00\pm0.04$ & $1.69\pm0.03$ & $1.73\pm0.06$\\
$v$ & 3 & $-2200\pm100$ & 0$^{f}$ & 0$^{f}$\\
\hline
$N_{\mathrm{H}}$ & 4 & -- & 22.00$^{f}$ & $21.29\pm0.08$ \\
$\log\xi$ & 4 & -- & $1.78\pm0.03$ & $1.39\pm0.05$\\
$v$ & 4 & --  & $-1730\pm50$ & $-1910\pm140$\\
\hline
$N_{\mathrm{OVII}}$ & & 17.81 & 18.11 & 17.95 \\
$N_{\mathrm{OVIII}}$ & & 18.14 & 18.20 & 17.97 \\
\hline
$\chi^2/dof$ & & 1.271 & 1.377 & 1.243 \\
 & & (8412.6 / 6620) & (16601.9/12053) & (14977.3 /
  12051) \\
\hline
\raggedright
$^{f}$ -fixed.
\end{tabular}
\end{table*}

\subsection{Relativistic emission line model}

\begin{figure}
\rotatebox{270}{
\resizebox{!}{\columnwidth}
{\includegraphics{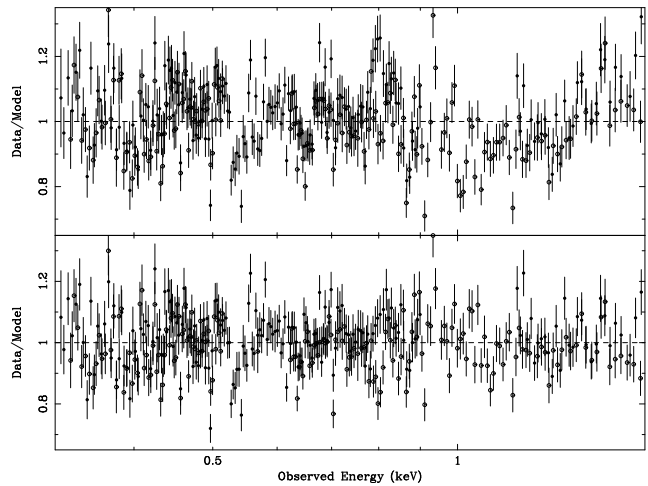}}}
\caption{  
  Ratio of data from RGS1 (\emph{filled circles}) and RGS2 (\emph{open
    circles}) to the best fitting REL models. Only data for orbit 301
  is shown here. (\emph{a}) Relativistically blurred emission lines
  and power-law (\emph{b}) Relativistically blurred emission lines and
  power-law with warm absorption.}
\label{fig:ratio_rel}
\end{figure}

\cite{brand-ray01} and \cite{sako01} argue that the drop in flux at
0.7~keV is due to the blue wing of a relativistically blurred OVIII
Ly$\alpha$ emission line. In this interpretation the warm absorber is
responsible for only narrow absorption features rather than any broad
ionization edges. In this section we fit this relativistic emission
line (REL) model to data from all three revolutions. In the original
model of \cite{brand-ray01} the model consists of a power-law and
emission lines due to CVI Ly$\alpha$ (0.368~keV), NVII Ly$\alpha$
(0.500~keV), and OVIII Ly$\alpha$ (0.654~keV) emitted by the
inner-part of an accretion disc around a rotating black hole. The
relativistic blurring is achieved with the {\tt LAOR} model
\citep{laor91}. Free parameters in the model are inner and outer
radius of the disc ($R_{\mathrm{in}}$ and $R_{\mathrm{out}}$), disc
inclination ($i$), emissivity $\epsilon$ as a function of radius $R$
which is parametrized as a power-law with index $q$ (i.e.
$\epsilon\propto R^{-q}$). Galactic absorption is also included. The
best fitting model produces a fit with $\chi^2 = 17457.8/12051$~\dof\ 
(see Fig.~\ref{fig:ratio_rel}a). The fit parameters of this model are
given in Table \ref{table_rel} (model 1). While the level of the
continuum is well reproduced, the region across the 0.7~keV drop is
poorly fitted since no FeI L-shell absorption is included. Adding FeI
to account for the L$_{2,3}$ edge clearly seen in the data
($N_{\mathrm{FeI}}=10^{17.2}\mathrm{cm}^{-2}$) produces a fit with
$\chi^2 = 17534.6/12051$~\dof. See Table \ref{table_rel} (model 2) for
fit parameters. The effects of warm absorption are now included by
adding absorption by OVI-OVIII, FeXVII-XXIV and NeIX-X in two velocity
components, one with an outflow velocity of $\sim2000$~\kmps and the
other at rest in the frame of the source as was observed by
\cite{sako01}. Also included is a broad Gaussian absorption trough at
$\sim$0.75~keV to represent any Fe UTA.  Allowing the columns of these
species and FeI to be free parameters improves the fit ($\chi^2 =
17579.4/12022$; see Fig.~\ref{fig:ratio_rel}b). Table \ref{table_rel}
(model 3) gives the fit parameters for this model. The OVIII column
density of $10^{18.4}\mathrm{cm}^{-2}$ produces an edge at 0.871~keV
with $\tau\sim0.24$. The best-fitting column of FeI required is
insufficient to account for the fine structure seen at 0.7~keV (see
Fig.~\ref{fig:ratio_detail}b) and the spectral index of the power-law
is $\sim$1.5, much lower than the $\sim$2.0 index found for the EPIC
data \citep{fabian02}. The equivalent widths of the three emission
lines are $\sim$34~eV for CVI Ly$\alpha$, $\sim$38~eV for NVII
Ly$\alpha$ and $\sim$220~eV for OVIII Ly$\alpha$.

Under the REL model of \cite{sako01} the 1.0--1.7~keV region consists
of a power-law modified by narrow absorption features. Here we fit
this region to a power-law producing a best-fitting spectral index of
$\Gamma=0.93\pm0.04$.  Even with a column of FeI of
$10^{17.2}\mathrm{cm}^{-2}$ the best-fitting spectral index is
$1.10\pm0.04$. In order to match the continuum here with the softer
power-law found at higher energies the continuum is required to break
at around 2~keV from a hard power-law below the break to a softer one
above. No mechanism is known that can create such a break.

\begin{table*}
\centering
\caption{Results of fits of the REL model to all three revolutions
  of RGS data. Equivalent widths of the emission lines ($EW$) are
  given in eV, disc radii ($R_{\mathrm{in}}$ and $R_{\mathrm{out}}$) are given in units of
  $r_{\rm g}$, inclination ($i$) is given in degrees and column
  density ($N_{\mathrm{H}}$) is given in units of $\log$\pscm.\label{table_rel}}
\begin{tabular}{cccc}
\hline
model & 1 & 2 & 3 \\
\hline\hline
$\Gamma$ & $1.46\pm0.02$ & $1.75\pm0.03$ & $1.50\pm0.04$ \\ 
$EW_{\mathrm{CVI}}$ & $35.2\pm12.7$ & $27.6\pm5.2$ & $34.1\pm10.0$ \\ 
$EW_{\mathrm{NVII}}$ & $59.6\pm3.6$ & $16.2\pm3.8$ & $38.4\pm15.5$ \\ 
$EW_{\mathrm{OVIII}}$ & $417.3\pm7$ & $167.7\pm4.2$ & $218.3\pm9.3$ \\ 
\hline
$q$ & $3.81\pm0.02$ & $4.68\pm0.07$ & $3.79\pm0.08$ \\ 
$R_{\mathrm{in}}$ & $2.42\pm0.01$ & $1.90\pm0.27$ & $2.11\pm0.02$ \\ 
$R_{\mathrm{out}}$ & $185\pm85$ & $13.69\pm3.60$ & $11.2\pm0.3$ \\ 
$i$ & $37.3\pm0.15$ & $45.1\pm0.18$ & $40.0\pm0.2$ \\ 
\hline
$N_{\mathrm{H}}$(FeI) & -- & 17.2$^{f}$ & $16.78\pm0.03$ \\ 
$N_{\mathrm{H}}$(OVII) & -- & -- & $17.32\pm0.08$ \\ 
$N_{\mathrm{H}}$(OVIII) & -- & -- & $18.39\pm0.03$ \\ 
\hline
$\chi^2/dof$ & 1.449 & 1.455 & 1.310 \\ 
 & (17457.8/12051) & (17534.6/12051) & (15749.4/12022)\\ 
\hline
\raggedright
$^{f}$ -fixed.
\end{tabular}
\end{table*}

\subsection{Combined DWA/REL model}
  
Finally the DWA model was combined with the REL model and fitted to
the data. It was found that the best fitting model obtained from the
fit was strongly dependant on the starting position of the fit. If the
REL model with zero line normalisation was added to the best fitting
DWA model (model 2) then only very small emission lines were required.
When the DWA model was added to the best fitting REL model (model 1)
then very little warm absorption was required. It would appear that
two separate solutions exist; one with a large DWA and small REL and
the other with small DWA and large REL.

\subsection{DWA/REL fitting summary}

The REL model by itself is a poor fit and requires the addition of a
DWA. Even then some parameters are unphysical. The simple DWA model,
acting on a power-law continuum, probably requires structure in the
soft X-ray continuum. The likely best-fitting model is a DWA acting on
a continuum with some emission structures.

\section{Empirical absorption model}

\label{empirical_model}

The principal difficulty in fitting the soft X-ray spectra of absorbed
Seyfert 1 AGN is that the shape and level of the continuum is not
known, partly because broad emission lines can mimic absorption edges.
In this section we attempt to circumvent this problem using the known
spectral variability of \mcg.

The flux in the Fe K$\alpha$ line at $\sim$6.4~keV is observed to vary
little while the flux in the continuum is highly variable
\citep*{lee00,vaughan01,lee02,shih02,fabian02}. If a spectrum
extracted from a period of low continuum flux is subtracted from a
spectrum extracted from a period of higher flux, the resulting
spectrum (the difference spectrum) can be adequately fit by just a
power-law above $\sim$2~keV \citep{fabian02}. It would therefore
appear that above $\sim$2~keV the emission consists of a rapidly
variable power-law with constant Fe K$\alpha$ emission. On this basis
\cite{fabian03} introduce a model for the emission consisting of a
rapidly variable Power-Law Component (PLC) and an almost constant
Reflection Dominated Component (RDC). Since the RDC changes little,
subtracting a low flux spectrum from a high flux spectrum effectively
removes it.

\begin{figure}
\rotatebox{270}{
\resizebox{!}{\columnwidth}
{\includegraphics{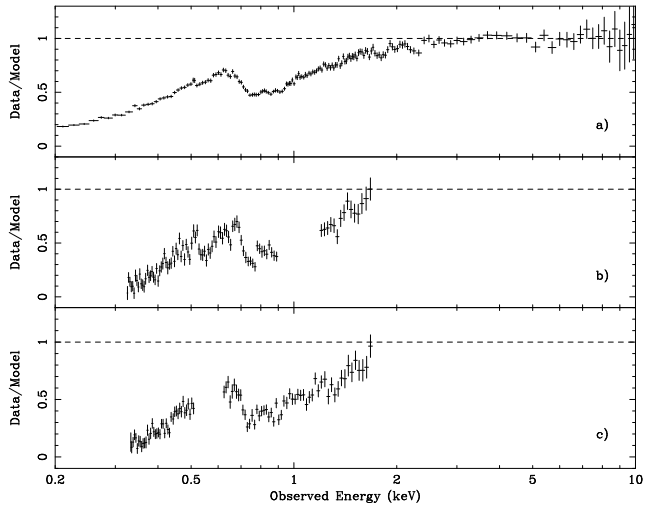}}}
\caption{
  (\emph{a}) Data/model ratio plot from fitting the pn difference
  spectrum to a power-law in the range 3--10~keV, excluding Galactic
  absorption. (\emph{b}) Data/model ratio plot for the RGS1 difference
  spectrum compared to the model from (\emph{a}).  (\emph{c}) Same as
  (\emph{b}) except for the RGS2 difference spectrum. The gaps in the
  RGS data are due to nonfunctional chips in the detectors.}
\label{fig:combined_absorption}
\end{figure}

Data from revolution 303 were used to examine this variability for the
RGS data since the flux level varied considerably during this period
(see \citealt{vaughan03} for light curves). Two pn spectra were
extracted, one from the first 80~ks when the mean flux level was high
and one from the last 40~ks when the mean flux level was low. The low
flux spectrum was subtracted from the high flux spectrum and the
resulting difference spectrum was found to be adequately fitted by a
power-law with $\Gamma\sim2.2$ in the range 3--10~keV. The ratio of
data to best fitting power-law is shown in
Fig.~\ref{fig:combined_absorption}a. Assuming that the variability
model applies to any reflection components below 3~keV, the underlying
continuum in the pn difference spectrum is just a power-law and any
deviations that are seen in Fig.~\ref{fig:combined_absorption}a are
due to absorption. These deviations from the high energy power-law
therefore represent the absorption function of the source, including
both intrinsic and Galactic absorption.

\begin{figure}
\rotatebox{270}{
\resizebox{!}{\columnwidth}
{\includegraphics{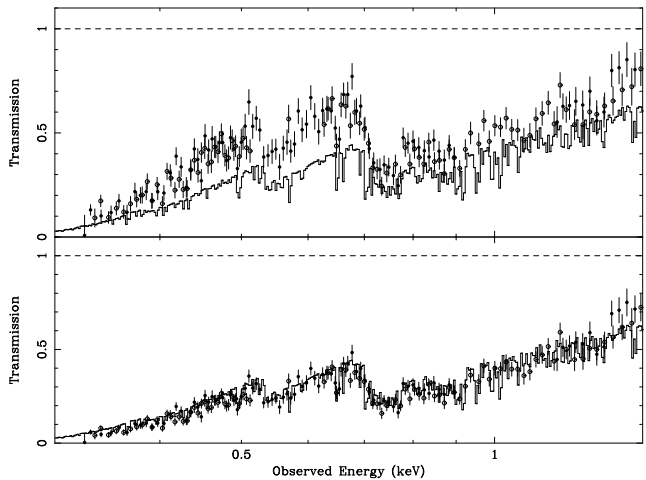}}}
\caption{Empirical absorption function derived from RGS1 (\emph{filled
    circles}) and RGS2 (\emph{open circles}) compared to the best
  fitting DWA model absorption model (\emph{solid line}). (\emph{a})
  Absorption function derived from the pn difference spectrum.
  (\emph{b}) Absorption function derived from a broken power-law.}
\label{fig:abs_func_fp}
\end{figure}

\begin{figure}
\rotatebox{270}{
\resizebox{!}{\columnwidth}
{\includegraphics{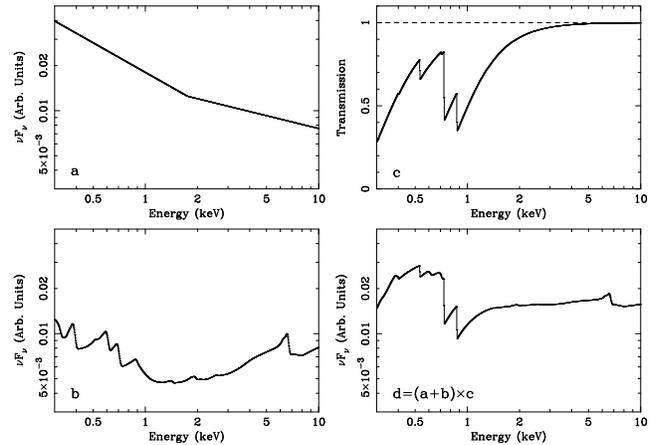}}}
\caption{Model components present in spectra. (\emph{a}) Power-Law
  Component (PLC). (\emph{b}) Reflection Dominated Component (RDC)
  showing strong Fe K$\alpha$ emission and weaker soft X-ray emission
  lines. (\emph{c}) Dusty Warm Absorber (DWA) showing strong OVII and
  OVIII edges. (\emph{d}) Combination of all model components. N.B.
  This diagram is purely schematic and does not represent the actual
  model components.}
\label{fig:plot_new}
\end{figure}

Next RGS spectra for the first 80~ks and last 40~ks of revolution 303
were extracted. The RGS spectra from the low flux period was
subtracted from the RGS spectra from the high flux period for each
instrument to leave the PLC modified by absorption. These RGS
difference spectra were then divided by a power-law, with the spectral
index and normalization fixed to the values derived from the pn
difference spectrum, to produce the RGS absorption functions in
Fig.~\ref{fig:combined_absorption}b and c. The OI edge at
$\sim$0.54~keV and the FeI/OVII edge at $\sim$0.7~keV can clearly be
seen. This absorption function is compared to the best-fitting DWA
absorption model from section \ref{DWA_model} in
Fig.~\ref{fig:abs_func_fp}a where the best-fitting DWA model is seen
to be more absorbed than the absorption function inferred from the pn
difference spectrum. This would result if the varying PLC is a broken
power-law with a break at $\sim1-2$~keV and a softer spectrum below
this break than that inferred from the pn difference spectrum in the
3--10~keV range.  Fig.~\ref{fig:abs_func_fp}b compares the empirical
absorption function obtained by assuming the PLC to be a broken
power-law with slopes $\Gamma_1=2.66$, $\Gamma_2=2.29$ and break
energy $E_{\mathrm{br}}=1.75$~keV, with the best-fitting DWA model.
Good agreement is found.  Also if the best fitting DWA model is
applied to the pn difference spectrum a broken power-law for the
underlying continuum produces a good fit to the data.
Fig.~\ref{fig:plot_new} shows the different model components present
in the spectra.

\begin{figure}
\rotatebox{270}{
\resizebox{!}{\columnwidth}
{\includegraphics{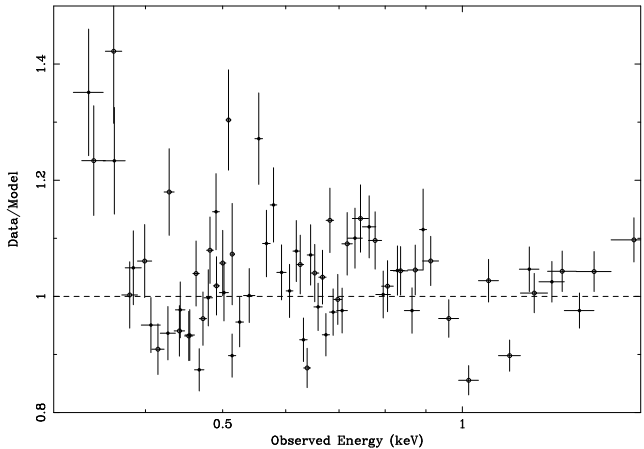}}}
\caption{
  Combined ratio of the absorption corrected data from RGS1
  (\emph{filled circles}) and RGS2 (\emph{open circles}) to the
  best-fitting power-law for all three orbits.}
\label{fig:abs_corr_comb}
\end{figure}

We proceed by assuming that the absorption function for the soft X-ray
spectrum is the pattern of deviations in the RGS difference spectrum
from a broken power-law. This will be the case if the variable part of
the spectrum is approximately a broken power-law continuum.
Specifically this means that any additional emission features do not
vary with the continuum, which would mean they contribute to the
reflection dominated component of \cite{fabian03}. The assumption is
supported by the fact that the RMS variability spectrum shows no sharp
emission or absorption features below 1~keV \citep{fabian02} and is
well represented by the above model. The spectral indices and break
energy of the broken power-law are assumed to be fixed while the
normalization varies. Such a continuum is observed in many other
Seyfert 1 X-ray spectra. The RGS difference spectra are divided by the
broken power-law to form the RGS absorption function.  Now that the
absorption function has been derived for the soft X-ray band, its
effects can be removed from the RGS data for the whole observation and
the RDC in the soft-band can be examined. RGS data for all three
revolutions were divided by the absorption function to correct for
both Galactic and warm absorption. The spectra for each revolution
were then fitted to a power-law and the ratio of data to best fitting
model for each revolution combined to produce
Fig.~\ref{fig:abs_corr_comb}. The photon index of the best fitting
power-law for the absorption corrected spectra was $\sim$2.8, which is
steeper than that derived for the high energy spectral index of the
PLC of $\sim$2.3, re-enforcing the idea that there may be a soft
excess below 2~keV, similar to the variable two component power-law in
Ton S180 \citep{vaughan02}. The residuals also show an excess of
emission below 0.4~keV and a trough at $\sim$1~keV. Any features
around the OVII (0.5740~keV) or OVIII (0.6536~keV) emission lines are
constrained to be less than $\sim$20 per cent of the continuum.
Provided the PLC continuum is a power-law in the RGS passband, the
shape of the residuals in Fig.~\ref{fig:abs_corr_comb} are independent
of the spectral index and normalization assumed for that power-law.
Note the resemblance to the residuals found earlier
(Fig.~\ref{fig:ratio_dwa}a).

The whole emission complex between 0.45 and 0.9~keV was earlier
modelled as a broad Gaussian (section \ref{DWA_model}). We now fit it
in the absorption corrected spectra with a power-law and emission
lines centred on the OVII $1s^2-1s2p$ and OVIII Ly$\alpha$ line
energies. Fits with both intermediate width Gaussian and
relativistically-blurred lines produce upper limits on the equivalent
width of OVII and OVIII emission of $\sim30$~eV. The whole complex
could be due to OVII, OVIII, FeL and other emission expected in the
reflection spectrum \citep{ballantyne02} together with further
absorption structure.  Detailed modelling of this emission will be
carried out in later work.

\section{Discussion}

\label{discussion}

The high signal-to-noise of our 320~ks observation has allowed the
unambiguous detection of FeI L-shell absorption through the fine
structure at the L$_{2,3}$ edge. This has been observed before in the
ISM \citep{paerels01} and in Cygnus X-1 \citep{schulz02}. It combines
with the resonance and edge absorption from OVII and OVIII to explain
the drop in flux at 0.7~keV \citep{lee01} and remove the
$\sim$16,000~\kmps discrepancy in position between the observed drop
and that expected from the OVII K-shell edge reported in
\cite{brand-ray01}.  

In comparing the DWA and REL models to the data, both produce similar
values of $\chi^2$, although all models are formally unacceptable.
However the normalizations of the three broad emission lines in the
REL model are free parameters, so the level of the continuum in each
section of the spectra can be easily fit. The equivalent widths of
these lines are also much larger than has been obtained with
self-consistent models of ionized disc reflection
\citep{ballantyne02}. As has been noted in Section
\ref{spectral_fitting}, the best-fitting REL model requires
insufficient FeI to fit properly the fine structure at 0.7~keV and
produces an unrealistically flat spectral index for the underlying
power-law. The DWA model in contrast uses only a broken power-law for
the underlying continuum. Addition of a reflection component, with
emission features around the 10--20 per cent level (Fig.
\ref{fig:abs_corr_comb}), may well explain some of the remaining
residuals (see for example the model in \citealt{fabian03}).  The DWA
model also assumes Solar abundances and a turbulent velocity of 100
\kmps.  Relaxing these two assumptions should improve the fit further.
These improvements will be explored in a later paper.

In Section \ref{empirical_model} the pn and RGS difference spectra are
used to determine the presence of additional emission components.
This method reveals a steep increase in flux below 0.4~keV and a
trough at 1~keV. This suggests that an extra emission component is
present below 0.4~keV which, combined with increased low ionization
parameter warm absorption, could explain the low energy residuals from
the DWA model. This extra emission component may be the high energy
tail of the quasi-blackbody emission emitted from the accretion disc.
These methods assume the absorption present is non-varying, whereas an
anti-correlation of the optical depth of an absorption feature with
ionizing flux would produce a drop in the absorption corrected spectra
where that absorption was. Such variability was observed by
\cite{otani96} for absorption around 1~keV.  The residuals in the DWA
fit at 1~keV may be due to a warm absorber zone with this sort of
variability which would explain the trough seen at $\sim1$~keV in the
absorption corrected spectra. The fact that no large peak in the
absorption spectra is observed in the 0.6--0.7~keV region suggests
that any OVIII emission, which does not vary exactly with the
continuum, is at a level not greater than $\sim$10--20 per cent of the
continuum. This is much smaller than in the basic REL model.

In summary, we have found that detailed absorption modelling of the
average RGS spectra agrees with an analysis based on variability. Both
indicate the dominant presence of dusty warm absorber. Residuals from
the modelling show that the underlying continuum has some complex
emission structures in the 0.4--0.9~keV band of amplitude consistent
with theoretical predictions for an ionized reflector.

\section*{Acknowledgements}

Based on observations obtained with \xmm, an ESA science mission with
instruments and contributions directly funded by ESA Member States and
the USA (NASA). AKT acknowledges support from PPARC. ACF thanks the
Royal Society for support. JCL thanks the Chandra fellowship for
support. This was provided by NASA through the Chandra Postdoctoral
Fellowship Award number PF2-30023 issued by the Chandra X-ray
Observatory Center, which is operated by SAO for and on behalf of NASA
under contract NAS8-39073. We thank Gary Ferland for help with
\cloudy\, Jeff Kortright for useful information on the FeI L edge
profile and an anonymous referee for helpful comments.

\bibliographystyle{mnras}                       %% MNRAS
\bibliography{mn-jour,turner_july14}

\end{document}